\documentclass[12pt,preprint]{aastex}
\usepackage{amsmath}
\newcommand{\be}{\begin{equation}}
\newcommand{\ee}{\end{equation}}

\begin{document}

\title{Quantifying parameter errors due to the peculiar velocities of type Ia supernovae}

\author{R. Ali Vanderveld}
\affil{Jet Propulsion Laboratory, California Institute of Technology, 4800 Oak Grove Drive, M/S 169-327, Pasadena, CA 91109, USA}
\email{rav@caltech.edu}

\begin{abstract}

The fitting of the observed redshifts and magnitudes of type Ia supernovae to what we would see in homogeneous cosmological models has led to constraints on cosmological parameters.  However, in doing such fits it is assumed that the sampled supernovae are moving with the Hubble flow, i.e.~that their peculiar velocities are zero.  In reality, peculiar velocities will modify supernova data in a way that can impact best-fit cosmological parameters.  We theoretically quantify this effect in the nonlinear regime with a Monte-Carlo analysis, using data from semi-analytic galaxy catalogs that are built from the Millennium N-body simulation.  We find scaling relations for the errors in best-fit parameters resulting solely from peculiar velocities, as a function of the total number of sources in a supernova survey $N$ and its maximum redshift $z_{\rm max}$.  For low redshift surveys, we find that these errors can be of the same order of magnitude as the errors due to an intrinsic magnitude scatter of $0.1$ mag.  For a survey with $N=2000$ and $z_{\rm max}=1.7$, we estimate that the expected peculiar velocity-induced errors in the best-fit cosmological constant density and equation of state can be $\sigma_{\Lambda}\approx 0.009$ and $\sigma_{w}\approx 0.01$, respectively, which are subdominant to the errors due to the intrinsic scatter.  We further find that throwing away supernova data below a redshift $z\approx 0.01-0.02$ can reduce the combined error, due to peculiar velocities and the intrinsic scatter, but by only about $10\%$.
\end{abstract}

\section{Introduction}

Type Ia supernovae (SNe) at a given redshift appear to be dimmer than one would expect in a homogeneous Friedmann-Robertson-Walker (FRW) Universe containing only pressureless matter and with gravity governed by general relativity 
\citep{Riess1,  Perlmutter, Knop,  Riess2}.  
This implies that the expansion rate of the Universe is accelerating.  Furthermore, if we restrict ourselves to fitting the observed redshifts and magnitudes of type Ia SNe to the predictions of FRW models, we find that the data appear to be best fit by a flat model with a matter density $\Omega_M\approx 0.25$ and a cosmological constant density $\Omega_{\Lambda}\approx 0.75$.  In this model, the Universe is presently undergoing accelerated expansion because of the current phase of dark energy domination.

There have been recent claims that we may not be justified in fitting SN data with FRW models, and thus doing so may have led to an incorrect assessment of the composition and behavior of the Universe; for a review, see 
Celerier (2007)
and references therein.  Of course, even if the Universe is homogeneous on large scales, local large scale structure does undoubtedly perturb the redshifts and apparent magnitudes of type Ia SNe.  So the question at hand is: Can such perturbations have a non-negligible impact on the inferences that we draw from these data? It has been suggested that the answer could be ``yes", with the largest effects coming from peculiar velocities and weak lensing 
\citep{HG,CC}. 
We will focus here on peculiar velocities, whereas 
Sarkar et al.~(2007)
performs a complementary analysis with the focus on lensing.  Although it is really the total combined effect of inhomogeneity that is gauge invariant and observable, we are nonetheless allowed to look at the peculiar velocity effect alone here in the Newtonian regime.

The goal of this paper is to theoretically quantify the parameter errors that result from the peculiar velocities of type Ia SNe in the most realistic framework possible.  We will thus make use of N-body simulation data, making this the first theoretical study of this effect that robustly takes into account not only correlated bulk flows, but also fully nonlinear (Newtonian) structure formation, both of which are thought to enhance the effect.  

We will find that our results are in accordance with recent estimates from actual SN data.  A key benefit of our theoretical approach, however, is that we can now estimate the size of the effect that we will expect for future SN surveys, such as SNAP \footnote{http://snap.lbl.gov} for which we find the peculiar velocity-induced error in a constant dark energy equation of state to be at the $1\%$ level.  This is subdominant to the $\sim 5\%$ error, that we would find with the same survey and model fitting, due to a $0.1$ mag intrinsic scatter. We will also address the question of whether it is beneficial to throw away very low redshift ($z\lesssim 0.02$) SN data when fitting to a model, as such data points will be the most affected by these errors; we find that such a practice could lead to error reductions of $10\%$ at most, for low redshift surveys.  These issues are of great importance at a time when the dark energy problem is at the forefront of modern science, and when we are planning ground and space-based facilities to gather SN data.

\section{Peculiar Velocities}

Now we will briefly review the effect of peculiar velocities on type Ia SN data, wherein we measure SN redshifts $z$ and luminosity distances $D_L$ to produce a ``Hubble diagram."  To lowest order, peculiar velocities add an extra redshift. For an observer with a peculiar velocity ${\bf v}_o$ and a source with a peculiar velocity ${\bf v}_s$, an unperturbed redshift $\tilde{z}$, and an unperturbed luminosity distance $\tilde{D}_L$, the final (perturbed) redshift is
\begin{equation}
z=\tilde{z}+\left(1+\tilde{z}\right){\bf n}\cdot\left({\bf v}_s-{\bf v}_o\right)
\label{zcorrection}
\end{equation}
and the final luminosity distance is
\begin{equation}
D_L(z)=\left(1+2{\bf n}\cdot{\bf v}_s-{\bf n}\cdot{\bf v}_o\right)\tilde{D}_L(\tilde{z})~,
\label{dlcorrection}
\end{equation}
where ${\bf n}$ is a unit vector along the line of sight, pointing from observer to source; these are equivalent to Eqs. (11) and (13) of 
Hui \& Greene (2006).  
Note that we have set the speed of light $c=1$.  Although it is typically assumed that these corrections are negligible, they actually can become quite important for very nearby SNe, for which $z\lesssim (10-100)v/c$, where $v\equiv |{\bf v}|$.  Having one end of the Hubble diagram with such errors will have more significant consequences than one might naively expect when it comes time to fit the data to a model.  

A further complication arises because the peculiar velocities are correlated, and correlated errors will not decrease as fast as $\sqrt{N}$, where $N$ is the number of SNe in the sample. From 
Cooray \& Caldwell (2006), 
in the limit of large $N$, the final variance of some measured parameter will be
\begin{equation}
\sigma^2 \approx \sigma^2_{0}\left[\frac{1+\left(N-1\right)r^2}{N}\right] \approx \frac{\sigma^2_{0}}{N}+\sigma^2_{0}r^2,
\label{error}
\end{equation}
where $r^2$ is the ratio of the average off-diagonal covariance matrix element to the average diagonal covariance matrix element, and both $\sigma_0$ and $r$ depend on the maximum survey redshift $z_{\rm max}$.  Note that $0\leq r^2\leq 1$, with $r\rightarrow 0$ corresponding to no correlation and $r\rightarrow 1$ corresponding to perfect correlation. Therefore, the first term on the right hand side of equation (\ref{error}) corresponds to the uncorrelated Poissonian part of the error, and the second term corresponds to the correlated part. We also see that the error cannot be arbitrarily reduced simply by having larger $N$, as there is a limit of $\sigma\rightarrow \sigma_0 r$ for $N\rightarrow\infty$. 

\section{Method}

Although recent studies with real SN data 
\citep{NHC, GLS}
have pointed to peculiar velocities being an important effect, it is useful to quantify this in a very controlled theoretical framework.  For that purpose, we use publicly available semi-analytic galaxy catalogs created at MPA 
\citep{SAM}, 
which make use of the Millennium N-body Simulation 
\citep{MS}.  
The Millennium Simulation followed the evolution of $\sim 10^{10}$ dark matter particles, each with a mass $8.6\times 10^8h^{-1}M_{\odot}$, in a periodic box of size $500/h~{\rm Mpc}$.  This was carried out within the framework of the standard $\Lambda$CDM cosmological model.  We show a table of the parameters of the simulation and the resulting data in Table~\ref{params}.  After semi-analytic galaxy modeling, the resulting data file contains the present-day, $z=0$, positions and velocities of $9,925,229$ galaxies in the simulation box.

We use these data to find the size of the typical errors that one should expect as a result of peculiar velocities.  To this end, we produce mock SN catalogs by sampling their host galaxies randomly from the $9,925,229$ galaxies in the semi-analytic data. We create these catalogs for varying values for the number of SNe $N$ and the maximum catalog redshift $z_{\rm max}$.  With this method we not only explore the effect of sample size, but we also explore the effect of varying redshift coverage.  Extending this coverage to higher $z_{\rm max}$ should reduce the resulting errors.  Reducing the low-redshift coverage, by throwing away data below a minimum redshift $z_{\rm min}$, also reduces errors by excluding more of the SNe for which the peculiar velocity perturbation is large.  However, increasing $z_{\rm min}$ also means a reduced $N$, which will increase errors. A key issue then is whether or not it is advantageous to impose such a lower redshift cutoff.  

Note that we will be limited by the finite size $x_{\rm total}=500/h~{\rm Mpc}$ of the simulation box, which implies $z_{\rm max}\leq 0.1725$ for an observer placed at a corner of the box. However, we can simulate surveys that go out to higher $z_{\rm max}$ by assuming that high redshift SNe are negligibly correlated, and thus that their velocities can be randomly drawn from a Gaussian distribution.  In this way, we will compute the errors expected in a survey like SNAP, 
with $N=2000$ and $z_{\rm max}=1.7$.

For each choice of survey parameters, we randomly create $10,000$ independent mock SN catalogs.  For each catalog, we also choose a random corner in which the observer will reside so as to reduce the effects of cosmic variance. We place observers at corners in order to maximize the possible redshift coverage of our mock surveys, and this has the implication that our surveys all cover one octant of the sky, or $\sim 5,000$ square degrees.  Then, for each SN, we find its comoving distance $\chi$ from the observer using our knowledge of the position of the observer and of the SN's host galaxy.  The unperturbed redshift $\tilde{z}$ is found by numerically inverting the function
\begin{equation}
\chi=c\int_0^{\tilde{z}}\frac{dz'}{H(z')}=\frac{c}{H_0}\int_0^{\tilde{z}}\frac{dz'}{\sqrt{\Omega_M(1+z')^3+\Omega_{\Lambda}}},
\end{equation}
where $h\equiv H_0/(100~{\rm km/s/Mpc})=0.73$, $\Omega_{M}=0.25$, and $\Omega_{\Lambda}=0.75$ are the values used in the underlying N-body simulation. The unperturbed luminosity distance is then $\tilde{D}_L=(1+\tilde{z})\chi$ and the unperturbed distance modulus is 
\be
\tilde{m}=5\log_{10}\left(\frac{\tilde{D}_L}{{\rm Mpc}}\right)+25~.  
\ee
Next, the unperturbed redshifts and distance moduli are perturbed to linear order in the host galaxy peculiar velocities given in the semi-analytic data, while we assume that the data have already been corrected for the (known) peculiar velocities of the observers. Note that we are justified in perturbing linearly in velocity, as $(v/c)^2\ll v/c$ for all galaxies. Thus, for a SN with unperturbed data $(\tilde{z}_i,\tilde{m}_i)$ and a peculiar velocity along the line of sight $v_i$, the perturbed redshift is $z_i=\tilde{z}_i+(1+\tilde{z}_i)v_i$ and the perturbed distance modulus is $m_i=\tilde{m}_i+10v_i/\ln(10)$. 

Finally, the perturbed data $(z_i,m_i)$ are fit to FRW models with a standard least-squares minimization. There are three different models that we consider, which are all simplifications of a general model that has four parameters: the rescaled Hubble constant $h$, the matter density today $\Omega_{M}$, the dark energy density today $\Omega_{de}$, and the dark energy equation of state $w$. Note that for actual SN observations one must fit the data for the calibration magnitude $M$; for our purposes here we will ignore this complication.  In this model, the Hubble parameter as a function of redshift is
\begin{equation}
\begin{array}{c}
\displaystyle H(z)=H_0\Big[\Omega_{M}(1+z)^3+\Omega_{de}(1+z)^{3(1+w)}\\
\displaystyle +(1-\Omega_M-\Omega_{de})(1+z)^2\Big]^{1/2}~,
\end{array}
\end{equation}
where $H_0=h(100~{\rm km/s/Mpc})$. The three simplified FRW models to which we fit are: 
\begin{itemize}
\item[(i)] a flat model ($\Omega_{M}+\Omega_{de}=1$) with a cosmological constant ($w=-1$), where we fit to find $h$ and $\Omega_{M}$, 
\item[(ii)] a curved model with a cosmological constant where we fit to find $h$, $\Omega_M$, and $\Omega_{\Lambda}\equiv\Omega_{de}$, and 
\item[(iii)] a flat model where we fit to find $h$, $\Omega_M$, and $w$.  
\end{itemize}
These models are summarized in Table~\ref{fits}.  For each model, we look at the statistical distributions of the best-fit parameters of our $10,000$ realizations.  We find that the distributions are roughly Gaussian with means that approximately equal the underlying simulation values.  We concern ourselves with the standard deviations, which are representative of the errors to be expected from peculiar velocities in each scenario.  

To check our results, we will use two auxiliary methods for estimating the parameter errors.  The first method involves generating ``synthetic" survey data, wherein we do not use the N-body data but instead we choose the velocities at random from a Gaussian distribution with the known variance $\langle v^2 \rangle$.  Then, proceeding as before, we will find the contribution to the error that is due solely to uncorrelated noise, i.e.~we will find the first term in equation (\ref{error}). Secondly, we can estimate Fisher matrices 
\citep{HG}, 
\begin{equation}
F_{\alpha\beta}=\sum_{i,j}\frac{\partial m_i}{\partial p_{\alpha}}\tilde{C}_{ij}^{-1}\frac{\partial m_j}{\partial p_{\beta}}~, 
\end{equation}
where $\tilde{C}_{ij}=\langle\delta m_i\delta m_j\rangle$, $p_{\alpha}=(h,\Omega_M,{\rm etc.})$, and the error in parameter $p_{\alpha}$ is $\sigma_{\alpha}=\sqrt{[F^{-1}]_{\alpha\alpha}}$.  The Fisher matrices provide a useful check of our code, and they also give us an analytic understanding of the scaling of the uncorrelated errors with $N$ and $z_{\rm max}$ for each case.  We will then fit the final variances as $\sigma_{\rm total}^2=\sigma_{\rm pois}^2+\sigma_{\rm corr}^2$, where $\sigma_{\rm pois}$ is the uncorrelated component of the error and $\sigma_{\rm corr}$ is the correlated component.  We expect that $\sigma_{\rm pois}$ scales as $1/\sqrt{N}$, and both $\sigma_{\rm pois}$ and $\sigma_{\rm corr}$ are power laws in $z_{\rm max}$ for $z_{\rm max}\ll 1$.  Hence, we expect errors of the form $\sigma_{\rm total}^2=Az_{\rm max}^a/N+Bz_{\rm max}^b$ for small $z_{\rm max}$; this expectation comes from the findings of 
Vanderveld et al.~(2007), 
Section VB.

\section{Results and Discussion}

For model (i), with $z_{\rm min}=0$, we show some of our results in Table~\ref{2param}.  Here $\sigma_h$ is the standard deviation of the errors in $h$ and $\sigma_M$ is the corresponding error in $\Omega_M$, where we are only considering the errors due to peculiar velocities.  These results are best fit by 
\be
\sigma_h^2\approx 2.5\times 10^{-4}\frac{z_{\rm max}^{-2}}{N}+2.6\times 10^{-12}z_{\rm max}^{-8.1}
\ee
and 
\be
\sigma_M^2\approx 0.0011\frac{z_{\rm max}^{-4}}{N}+5.8\times10^{-11}z_{\rm max}^{-9.5}~,
\ee
where the first term in each of these is the uncorrelated piece that matches what we find using results from our synthetic surveys.  This also matches what we expect from the Fisher matrix for this fit.  Furthermore, for $z_{\rm max}=0.1725$, we find that $r\approx 0.025$.  We find that this value of $r$ is consistent with the statistics of the galaxy velocities in the simulation box, where we estimated using $5,000$ random galaxies, 
\begin{equation}
r\approx \sqrt{\frac{\sum_{i\neq j}v_iv_j}{(N-1)\sum_iv_i^2}}\approx\sqrt{\frac{0.03125-0.00664}{4,999\times 0.00664}}\approx 0.027~.
\end{equation}

One further issue is whether or not it is advantageous to throw away low redshift data points.  To explore this, we add a random intrinsic scatter of $0.1$ mag to the SN magnitudes, in addition to the peculiar velocity error.  We then throw away all data points that have a final (observed) redshift below a cutoff $z_{\rm min}$.  The resulting errors are shown in Table~\ref{zmin}, for $N=500$ and $z_{\rm max}=0.1725$.  We thus find that the optimal minimum redshift is $z_{\rm min}\approx 0.02$, for which we find a $7\%$ reduction in total error.  We also find, for these survey parameters, that the error due to peculiar velocities is the same order of magnitude as the error due to the intrinsic scatter alone.

We give some of the results of fit (ii), for $z_{\rm min}=0$, in Table~\ref{3param}, where $\sigma_{\Lambda}$ is the error in $\Omega_{\Lambda}$.  These errors scale as 
\be
\sigma_h^2\approx 0.0035\frac{z_{\rm max}^{-2}}{N}+2.8\times 10^{-6}z_{\rm max}^{-2.2}~,
\ee
\be
\sigma_M^2\approx 0.43\frac{z_{\rm max}^{-6}}{N}+6.1\times 10^{-4}z_{\rm max}^{-6}~,
\ee
and 
\be
\sigma_{\Lambda}^2\approx 0.16\frac{z_{\rm max}^{-6}}{N}+0.0012z_{\rm max}^{-5.1}~.
\ee
This time we find $r=0.036$ when $z_{\rm max}=0.1725$, which is still consistent with our prior rough estimate from galaxy statistics.  By adding an intrinsic magnitude scatter of $0.1$ mag to a survey with $N=500$ and $z_{\rm max}=0.1725$, we now find the optimal minimum redshift to be $z_{\rm min}=0.01$, this time yielding a total error reduction of $9\%$.  We find once again that the error due to peculiar velocities is of the same order of magnitude as the error due to this scatter alone.

Riess et al.~(1998) 
fit to model (ii), for a survey that has a low-redshift sample of 34 SNe with $z<0.15$ and a high-redshift sample of 15 SNe with $0.16<z<0.62$ (one SN with $z=0.97$ was included in some of their analysis).  We (roughly) simulate this scenario by assuming that the SNe in the high-redshift sample are uniformly distributed and negligibly correlated, and thus their velocities can be randomly drawn from a Gaussian distribution.  With this method, we estimate the errors for 
Riess et al.~(1998)
to be $(\sigma_h,\sigma_M,\sigma_{\Lambda})\approx (0.009,0.7,0.6)$, which are about a factor of 3 smaller than the quoted total errors.  For a survey like that of 
Neill et al.~(2007), 
using data from the Supernova Legacy Survey 
\citep{SNLS}
with 44 SNe with $0.015<z<0.125$ and 71 SNe with $0.25<z<1$, we find that the expected errors are $(\sigma_h,\sigma_M,\sigma_{\Lambda})\approx (0.003,0.07,0.07)$, in accordance with the finding $\Delta\Omega_{\Lambda}= -0.04$ in that paper.  Furthermore, for a SNAP-like  
survey with 500 SNe with $z<0.17$ and 1500 SNe with $0.17<z<1.7$, we find $(\sigma_h,\sigma_M,\sigma_{\Lambda})\approx (0.0008,0.005,0.009)$.

For model (iii), proceeding as before, we give some of our results in Table~\ref{4param} and we find the following scaling relations reproduce the standard deviations when $z_{\rm min}=0$: 
\be
\sigma_h^2\approx 0.0041\frac{z_{\rm max}^{-2}}{N}+4.9\times10^{-6}z_{\rm max}^{-1.9}~, 
\ee
\be
\sigma_M^2\approx 0.040\frac{z_{\rm max}^{-6}}{N}+3.0\times10^{-5}z_{\rm max}^{-6.6}~,
\ee
and
\be
\sigma_w^2\approx 0.12\frac{z_{\rm max}^{-6}}{N}+3.9\times 10^{-4}z_{\rm max}^{-5.7}~, 
\ee
where $\sigma_w$ is the error in $w$.  We now find the optimal minimum redshift for this fit to be $z_{\rm min}=0.01$, with a total error reduction of $11\%$.  Furthermore, for the aforementioned SNAP-like survey, we estimate the peculiar velocity-induced error to be $(\sigma_h,\sigma_M,\sigma_{w})\approx (0.001,0.002,0.01)$.  

This error should be compared with the other sources of error in such a survey, such as the intrinsic SN magnitude spread and gravitational lensing.  Still for model (iii), we find that an intrinsic spread of $0.1$ mag produces the errors $(\sigma_h^i,\sigma_M^i,\sigma^i_{w})\approx (0.003,0.01,0.05)$, and from Sarkar et al.~(2007) we see that gravitational lensing produces a corresponding error of $\sigma_{w}^l\approx 0.006$.  This means that the noise caused by peculiar velocities is greater than that caused by lensing, but smaller than that caused by a $0.1$ mag intrinsic scatter.  Therefore total combined effect is dominated by the intrinsic scatter: $\sigma_{w}^{\rm total}\approx \sigma_{w}^i\approx 0.05$.

We can compare our results to those of Hui \& Greene (2006), who perform a linear-order analysis, to find that nonlinear corrections appear to be relatively unimportant, as one might expect.  For example, from Fig. 4 of Hui \& Greene (2006), which shows results for a similar SNAP-like survey that incorporates a low redshift sample, we see that the additional error on $w$ due to peculiar velocities is of order $\sigma_{w}\sim 0.01-0.02$.  Keeping in mind that our survey structures and model fits differ somewhat, this still shows that there is not a significant difference in the rough size of our results.  This is not to say, however, that the correlated nature of large scale structure does not play a crucial role in determining these errors, also as emphasized by Hui \& Greene (2006).

We now address some of the assumptions of the above analysis.  First, we considered only the case with a sky coverage of approximately $5,000$ square degrees.  This should not matter much, as 
Hui \& Greene (2006)
found that the area and geometry of a survey do not have a significant impact as long as the area is large enough and the geometry does not have any unusual features. Thus, our results should be representative of surveys with large areas for low redshifts; this would be true for any survey that incorporates the Nearby SN Factory data 
\citep{SNF}.  We also chose the galaxies at random within the simulation box, when in reality there are selection effects that modify the redshift distribution.  In our analysis, the total number of SNe as a function of redshift increased like $z^3$, where in reality this rises faster for low $z$ and then eventually drops off at $z\sim 1$.  This means that a real survey would have a larger fraction of its SNe at lower redshifts, which would amplify the peculiar velocity error computed here.

Furthermore, we ignored the velocities of SNe with respect to their host galaxies. Such internal velocities are typically not larger than the host galaxy velocities and they are not correlated between SNe in different galaxies 
\citep{HG}, 
meaning that they would only amplify somewhat the Poissonian pieces of the errors quoted above.  Another assumption in our analysis is that the Universe is roughly homogeneous on the scale of the simulation box, and thus we did not address what happens if there are very large scale bulk flows.

For each model, we found that excluding data points below a redshift of $z_{\rm min}\sim 0.01-0.02$ can improve the situation slightly, by reducing the total error by as much as $\sim 10\%$.  A far better alternative to cutting data could be to use local flow models to try to estimate, and subsequently correct for, peculiar velocities of local sources 
\citep{NHC}.  
An analysis of the potential benefits of such methods is the subject of future work.

\acknowledgements 
I thank Ira Wasserman, \'{E}anna Flanagan, Leonidas Moustakas, Jason Rhodes, and the anonymous referee for their helpful advice and review of the manuscript.  This work was carried out at the Jet Propulsion Laboratory, California Institute of Technology, under a contract with NASA.  The Millennium Run simulation used in this paper was carried out by the Virgo Supercomputing Consortium at the Computing Centre of the Max-Planck Society in Garching. The semi-analytic galaxy catalog is publicly available at http://www.mpa-garching.mpg.de/galform/agnpaper.

\clearpage
\begin{deluxetable}{ccc}
\tablecolumns{3}
\tablecaption{Parameters of the MPA semi-analytic galaxy catalogs produced from the Millennium N-body simulation.}
\tablehead{\colhead{Parameter} & \colhead{Definition} & \colhead{Value}}
\startdata
$\Omega_{M}$ & total matter density today & 0.25 \\
$\Omega_{b}$ & baryon density today & 0.045 \\
$\Omega_{\Lambda}$ & dark energy density today & 0.75 \\
$w$ & dark energy equation of state & -1 \\
$n$ & initial power spectrum slope & 1 \\
$\sigma_8$ & perturbation amplitude today & 0.9 \\
$N_{\rm total}$ & number of galaxies & $9,925,229$ \\
$x_{\rm total}$ & simulation box size & $500/h~{\rm Mpc}$ \\
$\langle v^2/c^2 \rangle$ & peculiar velocity variance & $4.271\times 10^{-6}$
\enddata
\label{params}
\end{deluxetable}

\begin{deluxetable}{ccccc}
\tablecolumns{5}
\tablecaption{Summary of the three models for our fits.}
\tablehead{\colhead{Model} & \colhead{$h$} & \colhead{$\Omega_{M}$} & \colhead{$\Omega_{de}$} & \colhead{$w$}}
\startdata
(i) &   Varied & Varied & 1-$\Omega_{M}$ & $-1$ \\
(ii) &  Varied & Varied & Varied & $-1$ \\
(iii) & Varied & Varied & 1-$\Omega_{M}$ &  Varied
\enddata
\label{fits}
\end{deluxetable}

\begin{deluxetable}{cccc}
\tablecolumns{4}
\tablecaption{Parameter errors for model (i) with $z_{\rm min}=0$.}
\tablehead{\colhead{$z_{\rm max}$} & \colhead{$N$} & \colhead{$\sigma_h$} & \colhead{$\sigma_M$}}
\startdata
0.1725 & 500  & 0.0043 & 0.057 \\
       & 750  & 0.0038 & 0.049 \\
       & 1000 & 0.0032 & 0.043 \\
0.15   & 500  & 0.0061 & 0.093 \\
0.13   &      & 0.0082 & 0.15
\enddata
\label{2param}
\end{deluxetable}

\begin{deluxetable}{ccc}
\tablecolumns{3}
\tablecaption{The $z_{\rm min}$ dependence for model (i), with $N=500$ and $z_{\rm max}=0.1725$, and an added intrinsic scatter of $0.1$ mag.}
\tablehead{\colhead{$z_{\rm min}$} & \colhead{$\sigma_h$} & \colhead{$\sigma_M$}}
\startdata
0    & 0.0073 & 0.099 \\
0.01 & 0.0069 & 0.094 \\
0.02 & 0.0068 & 0.093 \\
0.03 & 0.0070 & 0.095
\enddata
\label{zmin}
\end{deluxetable}

\begin{deluxetable}{ccccc}
\tablecolumns{5}
\tablecaption{Parameter errors for model (ii) with $z_{\rm min}=0$.}
\tablehead{\colhead{$z_{\rm max}$} & \colhead{$N$} & \colhead{$\sigma_h$} & \colhead{$\sigma_M$} & \colhead{$\sigma_{\Lambda}$}}
\startdata
0.1725 & 500  & 0.019 & 7.4 & 4.6 \\
       & 750  & 0.018 & 6.9 & 4.3 \\
       & 1000 & 0.016 & 6.4 & 4.0 \\
0.15   & 500  & 0.023 & 12 & 7.1 \\
0.13   &      & 0.026 & 18 & 10
\enddata
\label{3param}
\end{deluxetable}

\begin{deluxetable}{ccccc}
\tablecolumns{5}
\tablecaption{Parameter errors for model (iii) with $z_{\rm min}=0$.}
\tablehead{\colhead{$z_{\rm max}$} & \colhead{$N$} & \colhead{$\sigma_h$} & \colhead{$\sigma_M$} & \colhead{$\sigma_w$}}
\startdata
0.1725 & 500  & 0.021 & 2.6 & 4.4 \\
       & 750  & 0.018 & 2.3 & 3.9 \\
       & 1000 & 0.016 & 2.1 & 3.6 \\
0.15   & 500  & 0.023 & 3.7 & 6.1 \\
0.13   &      & 0.027 & 6.0 & 9.6
\enddata
\label{4param}
\end{deluxetable}

\end{document}